\documentclass[apjl]{emulateapj}
\accepted{22 October, 2012 }
\usepackage{times}

\slugcomment{}

\shorttitle{Radio K+A}
\shortauthors{Nielsen et al.}

\begin{document}


\title{The current star formation rate of K+A galaxies}


\author{Danielle M. Nielsen\altaffilmark{1,2}, 
Susan E. Ridgway\altaffilmark{3,2}, 
Roberto De Propris\altaffilmark{2} 
and Tomotsugu Goto\altaffilmark{4,5}}
\email{nielsen@astro.wisc.edu}

\altaffiltext{1}{University of Wisconsin, Madison, WI, USA}
\altaffiltext{2}{Cerro Tololo Inter-American Observatory, La Serena, Chile}
\altaffiltext{3}{National Optical Astronomy Observatory, 950 N. Cherry Avenue, Tucson, AZ 85719, USA}
\altaffiltext{4}{Institute for Astronomy, University of Hawaii, Honolulu, HI, USA}
\altaffiltext{5}{Subaru Telescope, Hilo, HI, USA}



\begin{abstract}

We derive the stacked 1.4 GHz flux from FIRST (Faint Images of the Radio Sky at Twenty Centimeters) survey 
for 811 K+A galaxies selected from the SDSS DR7. For these objects we find a mean flux density of $56\pm 9$~$\mu$Jy. 
A similar stack of radio-quiet white dwarfs yields an upper limit of 43 $\mu$Jy at a 5$\sigma$ significance to 
the flux in blank regions of the sky. This implies an average star formation rate of 
1.6~$\pm$~0.3 M$_{\odot}$ year$^{-1}$ for K+A galaxies. 
However the majority of the signal comes from $\sim$4\% of K+A fields that have aperture fluxes above 
the $5\sigma$ noise level of the FIRST survey. 
A stack of the remaining galaxies shows little residual flux consistent with an upper limit 
on star formation of 1.3~M$_{\odot}$ year$^{-1}$.
Even for a subset of 456 `young' (spectral ages $<$~250 Myr) K+A galaxies we find that the stacked 1.4 GHz flux 
is consistent with no current star formation. Our data suggest that the original starburst has been terminated 
in the majority of K+A galaxies, but that this may represent part of a duty cycle where a fraction of these 
galaxies may be active at a given moment with dusty starbursts and AGNs being present.

\end{abstract}





\section{Introduction}

K+A galaxies, also known as post-starburst galaxies (PSG), show spectra of a
strong Balmer absorption series superposed over a K-giant dominated spectrum
typical of early-type galaxies, implying the recent termination ($< 1$ Gyr) of 
a significant episode of star formation in an otherwise quiescent stellar population 
\citep{dressler83,couch87}. These objects may be the best examples of galaxies 
transitioning from the `blue cloud' (of star-forming objects) to the `red sequence', 
and have often been identified as possible progenitors of the lenticular population 
in clusters \citep{yang04,yang08}, rejuvenated early-type galaxies (e.g., \citealt{panuzzo07})
or the descendants of the blue galaxies observed in intermediate redshift clusters 
\citep{zabludoff96,poggianti99}.

The mechanism by which star formation is initiated and/or quenched in these
galaxies is still unclear. The majority of local K+A galaxies lie in the 
general field rather than clusters \citep{blake04,goto07,vergani10} and the processes 
that trigger and halt star formation may differ in these environments. At the 
same time, in the intermediate redshift clusters where K+A galaxies were originally
discovered, a significant fraction of current star formation is obscured by dust 
\citep{duc02,saintonge08,dressler09,haines09}. Dusty starbursts tend to have strong 
Balmer absorption but only weak OII emission as the young stars and HII regions that 
produce the emission lines tend to lie in regions of high extinction while the longer 
lived A-stars can migrate out of their native molecular cloud: this would be a viable
model to explain the K+A spectral class \citep{poggianti00}. Therefore it
is still unclear whether K+A galaxies are truly ``red and dead,'' although
the 24$\mu$m observations of \cite{dressler09} in Abell 851 imply that star 
formation has largely ceased in these objects.

Obscured star formation may also be detected via radio emission at 20 cm 
(1.4 GHz), which is produced by synchrotron radiation from high energy cosmic rays 
originating in supernova shells and yields a measure of the massive star formation rate \citep{condon92,
kennicutt98}. Although \cite{smail99} found evidence of recent star formation in 5 K+A galaxies, 
other studies find little evidence of ongoing star formation within small samples \citep{miller01,goto04}.
However, \cite{buyle06}
found that most K+A have substantial gas reservoirs, similar to those
of spirals of the same luminosity, and suggest that K+A galaxies may
be observed during a hiatus in an episodic star formation history or
that current star formation may be obscured. 

In this {\it Letter} we use a stack of radio images to further test the
current star formation rate and/or AGN activity in spectroscopically
selected K+A galaxies. A description of the data and the analysis is
provided in the next section, while we interpret and discuss our
results in section 3. We adopt the latest WMAP cosmological parameters
with $\Omega_{M}=0.27$, $\Omega_{\Lambda}=0.73$ and H$_0=71$ km~s$^{-1}$~Mpc$^{-1}$.

\section{FIRST radio stacking of the SDSS DR7 K+A sample}

We select a sample of  811 K+A galaxies from the Data Release 7 of the SDSS
\citep{york00,abazajian09} using an updated catalog of \cite{goto07}. 
Only objects classified as galaxies with a spectroscopic signal-to-noise $> 10$ per pixel are considered. 
The selection criteria of K+A galaxies 
are equivalent widths of H$\alpha > -3.0$ \AA, H$\delta > 5.0$ \AA\ 
and [OII] $> -2.5$ \AA, where emission lines are negative. Galaxies of redshift 
$0.35 < z < 0.37$ are excluded from the sample due to the $5577$\AA\ sky feature. 
The selected sample of K+A galaxies have redshifts ranging $0.02 < z < 0.4$. 

We then use data from the FIRST survey \citep{becker95} to derive a mean radio image of 
K+A galaxies. Typical detection limits for a single FIRST image are
$\sim$1 mJy, which, for the mean redshift from our SDSS DR7 sample of K+A galaxies, would
only allow us to detect star formation rates in excess of 30 M$_{\odot}$ yr$^{-1}$.

\begin{figure*}
\includegraphics[width=6in]{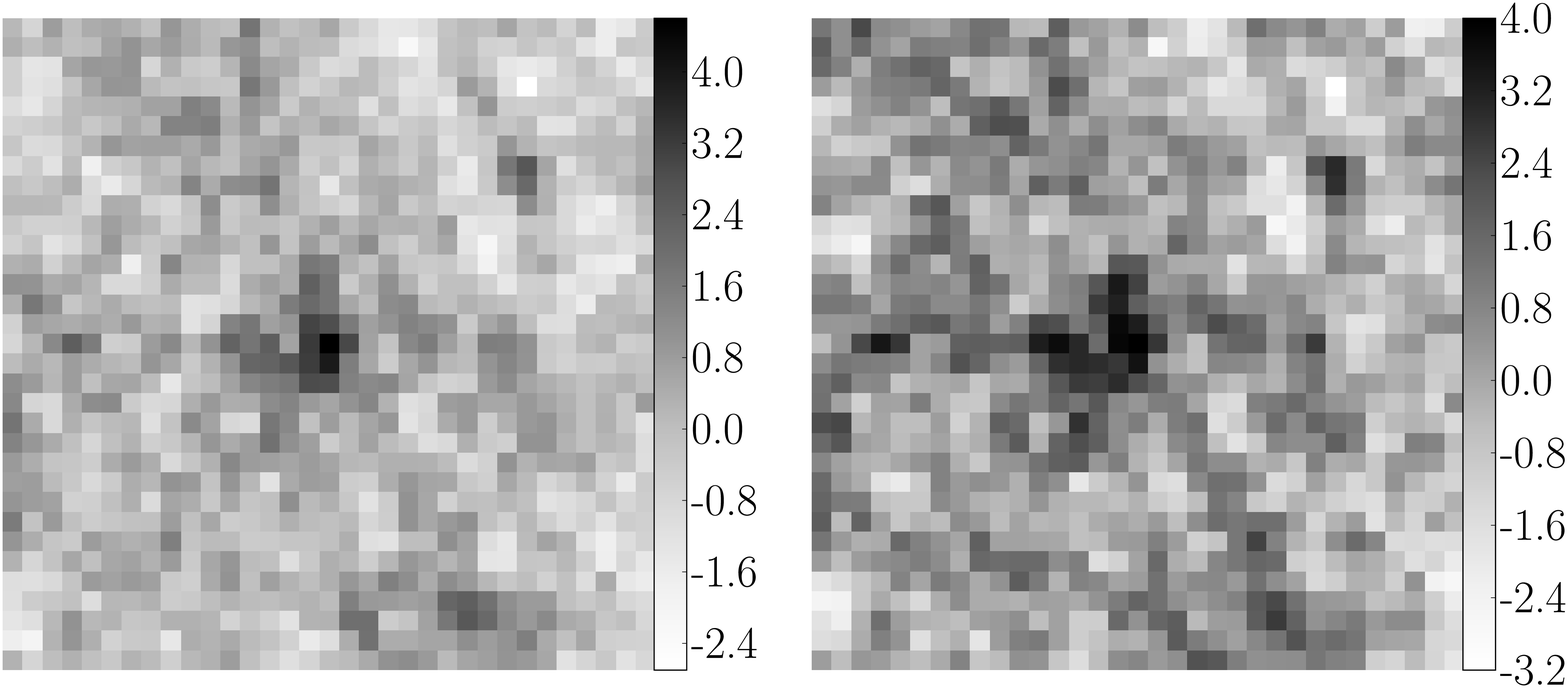}
\centering
\caption{{\sl Left:} The median stack of FIRST images of  811 K+A galaxies selected from the 
SDSS. The image is  $1'$ square with grayscale units in $\mu$Jy. 
{\sl Right: } The median stack of 427 K+A galaxies with less than than 250 Myr since the end of the 
starburst. The image is  $1'$ square with grayscale units in $\mu$Jy.}
\label{stackedFits}
\end{figure*}

For each of the  811 K+A galaxies covered by the survey we cut out a $1'$ square 
from the FIRST database. We use the median stacking method of \cite{white07} 	
to create the resulting image of  811 K+A galaxies shown in the left panel of Fig.~\ref{stackedFits}. 
We then measure the flux from the combined image in an aperture 
equivalent to three FIRST beams from which we derive an average K+A flux of $56\pm9$ $\mu$Jy.

To assess the significance of this result, we need to estimate the level of 
noise in the image. As FIRST is a survey, the image cutouts may not be 
fully cleaned of artifacts, such as side-lobes from distant radio sources.
To do this, we use a Monte Carlo simulation with a sample of known radio-quiet
objects, 8495 white dwarfs from the SDSS \citep{eisenstein06}. We create
10,000 stacks of 811 randomly selected white dwarfs (equivalent to the stack of 
K+A galaxies) and measure the flux in the same fashion to derive the mean flux 
from a supposedly radio-quiet sample.

\begin{figure}
\centering
\includegraphics[width=3.5in]{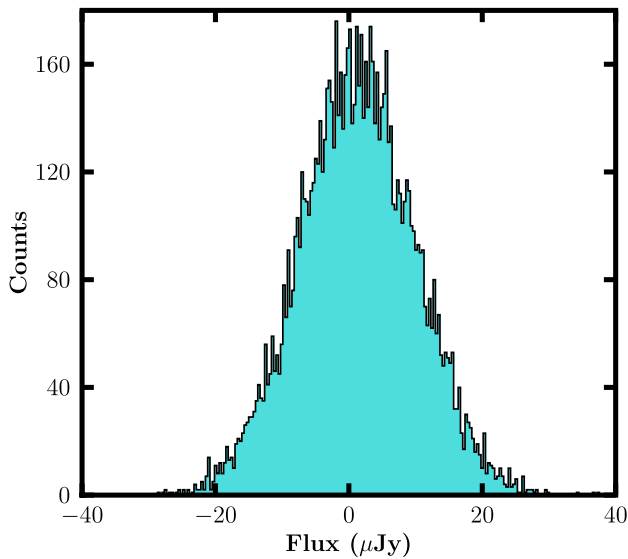}
\caption{Histogram of fluxes from Monte Carlo stacks of white dwarfs in the SDSS. 
Each stack is equivalent to the stack of K+A galaxies 
used in the analysis and contains 811 white dwarfs images. We find a 5$\sigma$ noise level of 43~$\mu$Jy. }
\label{wdHist}
\end{figure}

This procedure allows us to estimate the level of noise present in the 
stacked image and therefore to determine the significance of our detection for K+A 
galaxies. A histogram of the fluxes from our Monte Carlo simulation is shown 
in Fig.~\ref{wdHist}. Since we know that these sources are radio quiet, we can estimate a $5\sigma$ 
significance flux for sources to be considered real. We find this flux 
to be $43$~$\mu$Jy. Our K+A stack is found to have a significant detection above
the 5$\sigma$ noise level.

We can now compute the absolute luminosity at 1.4 GHz using our
chosen cosmology and the average redshift of the K+A galaxies ($z=0.14$).
\begin{equation}
L_{1.4 GHz}=4 \pi D^2_L S_{1.4 GHz} (1+z)^{\alpha} / (1+z),
\end{equation}
where $D_L$ is the luminosity distance, $S_{1.4 GHz}$ is the
flux density, $(1+z)^{\alpha}$ is the color correction and
$1/(1+z)$ the bandwidth correction \citep{morrison03}. We assume
that the radio emission is dominated by synchrotron radiation
such that $S \propto \nu^{-0.8}$ \citep{condon92}. This yields our
measurement of radio power, which we can
convert to a SFR using
\begin{equation}
\textrm{SFR (M}_{\odot} \textrm{ yr}^{-1}) = 5.9 \times 10^{-22} L_{1.4 GHz}(\textrm{W Hz}^{-1}),
\end{equation}
from \cite{yun01}, which assumes a Salpeter initial mass
function between 0.1 and 100 solar masses. This yields
a star formation rate of $1.6 \pm 0.3$ M$_{\odot}$ yr$^{-1}$ for our
average K+A galaxy at $<z>=0.14$.

However 79 of our sources are measured to have 1.4 GHz fluxes above the 3$\sigma$ noise level 
of the FIRST survey. The FIRST beam has an RMS of 0.15 mJy and we find 79 galaxies ($\sim$10\%) have 
aperture fluxes in excess of 3$\sigma$ (450 $\mu$Jy) within an aperture of 3 beamsizes, 31 of which have 
fluxes in excess of 5$\sigma$ (750 $\mu$Jy).

\begin{figure}
\centering
\includegraphics[width=3.5in]{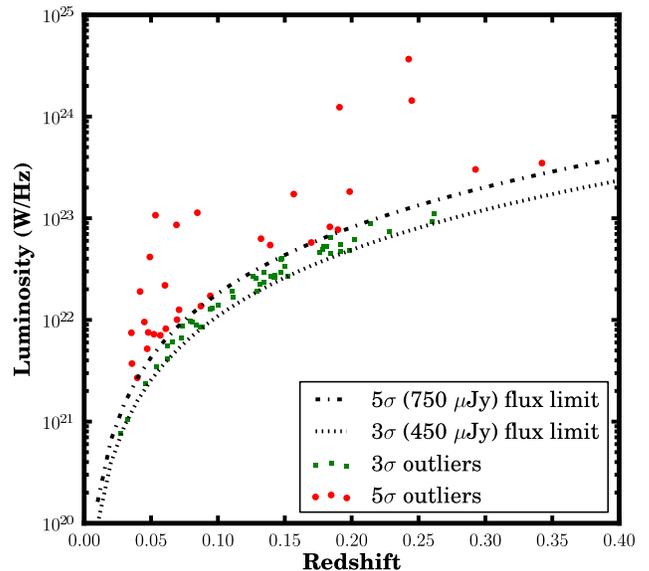}
\caption{We have plotted K+A galaxies from our sample with aperture fluxes in excess of 3$\sigma$ and 5$\sigma$ 
as green squares and red circles, respectively. The dotted and dot-dashed lines indicates the 3$\sigma$ and 5$\sigma$ 
FIRST survey limits, respectively, as a function of redshift.}
\label{luminosityZ}
\end{figure}

In Fig.~\ref{luminosityZ} we show the redshift--luminosity distribution of the targets with aperture fluxes
above the $3\sigma$ and $5\sigma$ limits. Visual inspection of these individual outlier frames have shown that for 
the 31 galaxies above the 5$\sigma$ limit, about 80\% look like clear individual detections 
(i.e., centrally concentrated, likely to be associated with our optical target, not in a frame that is 
significantly noisy), while for the 48 that are above the 3$\sigma$ limit but less than the 5$\sigma$ limit, 
only about 10\% of these are clear detections. Overall only about 4\% of our sample of 811 galaxies show evidence of 
significant ongoing radio activity, either from star formation or AGN activity. 

Removing all sources with measured fluxes in excess of 5$\sigma$, we create a 
subsample of 780 galaxies. Stacking this subsample yields a mean flux of 36 $\mu$Jy, 
which is well below the 5$\sigma$ detection limit of 47~$\mu$Jy found from a Monte Carlo 
simulation creating stacks of 780 white dwarfs. We can place an upper limit on the SFR of 
$1.3$~M$_{\odot}$~yr$^{-1}$ for this subsample with  $<z> = 0.14$.

When star formation ceases in a galaxy, we expect it to decrease exponentially
to low levels, rather than an abrupt truncation, unless some `catastrophic' event
 has removed all the available fuel or ionized the gas and 
prevented further star formation. Evidence for a rapid shutdown of star formation in 
these galaxies has been presented by \cite{brown09}. We therefore carry out the same analysis on
a subsample of `young' K+A galaxies and attempt to detect residual star formation
in these objects.

\begin{figure*}
\centering
\includegraphics[width=6.5in]{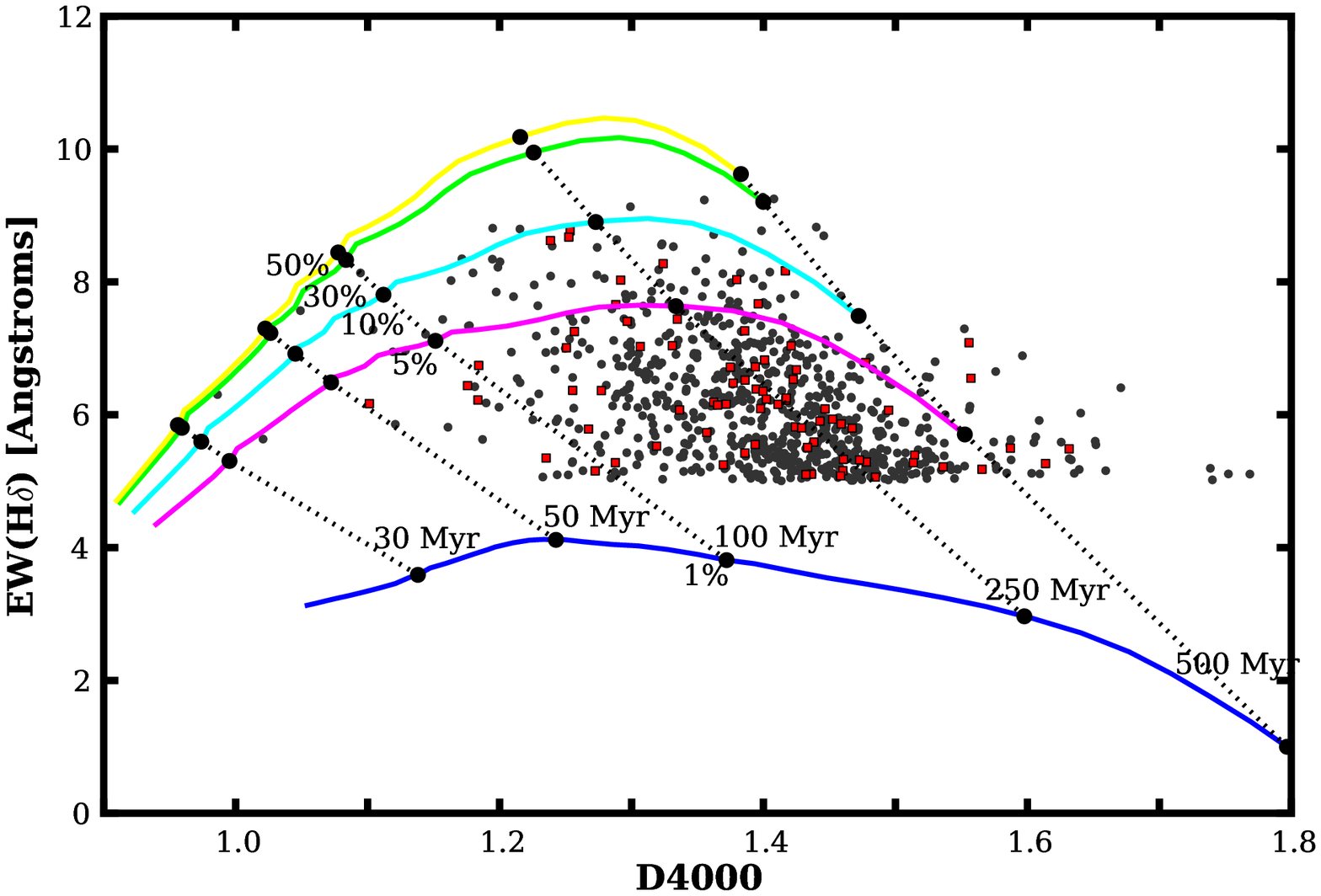}
\caption{Plot of H$\delta$ and D4000 for K+A galaxies from SDSS spectra and models. The gray circles and red squares 
are the measured values for K+A galaxies. The red squares indicate galaxies with significant 1.4 GHz detections. 
The solid lines are the models of an old 
10 Gyr exponentially decaying stellar population with a burst of 1\% (blue), 5\% (purple), 10\% (cyan), 
30\% (green) and 50\% (yellow) mass relative to the older stellar population. The dotted lines indicate the 
values of the spectral indices when observed 30, 50, 100, 250 and 500 Myr after the burst.}
\label{agePlot}
\end{figure*}

In Fig.~\ref{agePlot}, we plot H$\delta$ and D4000 from SDSS spectra and overplot models calculated with GALAXEV 
\citep{bruzual2003}. We plot the galaxies with significant 1.4 GHz detections as red squares, 
the rest as gray dots. We adopt a Salpeter initial mass function (IMF) and solar metallicity 
as initial conditions. The model galaxies evolved over 10 Gyr with an exponentially decreasing 
star formation rate ($\tau$ = 1 Gyr). At 10 Gyr, we added an instantaneous starburst (delta function) of mass 
1, 5, 10, 30 and 50\% (relative to the old stellar population) after which the SFR returns to zero. 
The dotted lines demarcate the values of the spectral indices observed 
30, 50, 100, 250 and 500 Myr after the burst. 
These are the same models used by \cite{yagi2006} and \cite{goto2008}.

We then select the 456 K+A galaxies with burst ages less than 250 Myr and an average
redshift of $<z> = 0.13$ and stack these galaxies 
in the same fashion as the complete sample. The stacked image is shown in the right panel of Fig.~\ref{stackedFits} 
from which we measure a mean flux of 61~$\mu$Jy. The Monte Carlo simulation with white dwarfs gives a 
5$\sigma$ detection limit of 59~$\mu$Jy. Our younger sample of K+A galaxies is found to be at the 
$5\sigma$ detection limit, giving a SFR of $1.5~\pm~0.3$~M$_{\odot}$ yr$^{-1}$. This is consistent with 
a rapid decline of star formation in these galaxies as shown by \cite{brown09}. 

\section{Discussion}

We have found an average star formation rate of only $1.6~\pm~0.3$~M$_{\odot}$ yr$^{-1}$ from
stacking radio observations of 811 {\sl general field} K+A galaxies in the local universe. 

However, much of this signal appears to originate from $\sim$4\% of active galaxies. Based on the
definition of \cite{sadler02}, which requires radio power above $10^{23}$ W Hz$^{-1}$ and the absence
of emission lines, this sample is approximately equally split between star forming galaxies and AGN (with the
latter more prominent at higher redshifts because of selection effects).

For the remainder of the sample, we find an upper limit on the SFR of 1.4 M$_{\odot}$ yr$^{-1}$. 
Even a subsample of spectroscopically young galaxies does not show a significant detection of residual star formation. 

Our results are in good agreement with previous work. \cite{goto04} did not detect evidence of star 
formation from VLA observations in  36 galaxies drawn from the SDSS DR1 \citep{goto03} 
but he was able to set an upper limit  of $< 15$ M$_\odot$ yr$^{-1}$ for 15 of the nearest K+A galaxies. 
\cite{miller01} found that only 2 out of 15 K+A galaxies in his sample show signs of 
obscured star formation. Nevertheless, \cite{melnick12} find that most K+A galaxies in this sample exhibit 
significant excess above the predicted stellar component at $\lambda>5$ $\mu$m from WISE data, suggesting 
that dust heated by some unknown source is present in these objects. 

\cite{buyle06} pointed out that K+A galaxies contain
significant amounts of gas and therefore the current quiescent star formation may
only be temporary, while \cite{poggianti00} proposed a model where star formation
continues in K+A galaxies but is hidden by dust. However, \cite{dressler09} detected
no 24 $\mu$m emission for K+A galaxies in Abell 851. 

Another possible contribution to the radio flux is that K+A galaxies may 
host an AGN. Most K+A galaxies have significant bulges \citep{yang08} and 
therefore should contain a supermassive central black hole (e.g., \citealt{ferrarese00}). 
In the general field, K+A galaxies are often involved in mergers and interactions 
\citep{blake04,goto05,yang08} and show inverted color gradients indicative of 
central star formation \citep{yang08}, which is expected if mergers drive gas to 
the center \citep{hernquist92}.  

\cite{liu07} find evidence of a weak AGN in a nearby K+A galaxy. 
However, \cite{shin11} find only a few AGNs in a subset of our sample and they argue that these 
active nuclei are not related to the quenching of the previous starburst. \cite{brown09} argue that 
AGN feedback is only important among the more massive K+A galaxies. \cite{vergani10} find no
evidence for AGN in their sample of COSMOS K+A galaxies using deep VLA and XMM imaging 
of the COSMOS field and only a small number of star forming objects, in agreement with our results. 
On the other hand \cite{georgakakis08} argued for a weak correlation between AGN and K+A galaxies 
in the EGS field. 

\cite{snyder11} present a model where K+A galaxies originate from gas-rich mergers and the duration of the 
Balmer-line strong phase is shorter than the commonly assumed $\sim$1 Gyr by about a factor of 3. In 
this framework the $\sim$4\% of `active' galaxies that we find may be the tail end of the distribution, 
while other objects are already heading towards the red sequence. In this case, this might yield a constraint 
on the activity timescale, assuming a 0.3 Gyr duration for the K+A phenomenon. If the 4\% of active galaxies 
represent the last gasp of activity before the AGN returns to a dormant state, the period of visible QSO 
activity may be estimated to be about 20 Myr, which is broadly consistent with current estimates 
(e.g., see review by \cite{martini04}). 

Another plausible explanation is that we are actually observing galaxies undergoing a series of starbursts 
and feedback episodes, as in the cycle postulated by \cite{hopkins08} to account for the tightness of the 
red sequence within a hierarchical formation framework. In which case we would be observing galaxies in 
the `on' and `off' phase of such a cycle. 

\cite{shioya02} present a model where truncated spirals evolve through the e(a) (galaxy with strong H$\delta$ 
and modest OII emission), a+k and finally k+a phase before settling on the red sequence. If we are observing 
galaxies at random along their spectrophotometric transformation, and given the $\sim0.3$ Gyr duration of the 
most visible k+a phase, the timescale for evolution of these galaxies on to the red sequence is long, $\sim7$ 
Gyr, which is consistent with the above simulations. Only low mass objects can then truly evolve on to the 
red sequence in reasonable times. 

A series of questions remain unanswered for future studies. Do K+A galaxies host on-going dusty 
starbursts? This appears unlikely in the light of our results, but the detection of powerful far 
infrared excesses by \cite{melnick12} points to the existence of extra sources of flux that are 
heavily obscured. What is the role of the AGN (if any) in modulating the rapid onset and decline of star 
formation? Do these objects contain cold gas and possibly re-initiate star formation? Future papers by our 
group will attempt to address some of these issues.

\acknowledgments

We thank Neal Miller for reading this paper and providing
helpful comments which improved our analysis.
We also thank Eric Wilcots and Mark Lacy for useful discussions.
We acknowledge the anonymous referee for providing constructive suggestions. 
This project was conducted in the framework of the CTIO REU Program,
which is supported by the National Science Foundation under grant
AST-0647604.



{\it Facilities:} \facility{Sloan, FIRST}.

\clearpage




\begin{thebibliography}{}
\bibitem[Abazajian et al.(2009)]{abazajian09}
Abazajan, K. et al. 2009, \apjs, 182, 543
\bibitem[Becker et al.(1995)]{becker95}
Becker, R. H., White, R. L., \& Helfand, D. J., 1995, \apj, 450, 559
\bibitem[Blake et al.(2004)]{blake04}
Blake, C. et al. 2004, \mnras, 355, 713
\bibitem[Brown et al.(2009)]{brown09}
Brown, M. I. J. et al. 2009, \apj, 703, 150
\bibitem[Bruzual \& Charlot (2003)]{bruzual2003}
Bruzual, G., \& Charlot, S. 2003, \mnras, 344, 1000 
\bibitem[Buyle et al.(2006)]{buyle06}
Buyle, P., Michielsen, S., De Rijcke, D., Pisano, D. J., Dejonghe, H. \& Freeman, K.
2006, \apj, 649, 163
\bibitem[Cohen et al.(2007)]{cohen07}
Cohen, A. S., Lane, W. M., Cotton, W. D., Kassim, N. E., Lazio, T. J. W., Perley, R. A., Condon, J. J. \& Erickson, W. C.
2007, \aj, 134, 1245
\bibitem[Condon(1992)]{condon92} Condon, J. J. 1992, \araa, 30, 575
\bibitem[Cordey(1987)]{cordey87} Cordey, R. A. 1987, \mnras, 227, 695
\bibitem[Couch \& Sharples(1987)]{couch87} 
Couch, W. J. \& Sharples, R. M. 1987, \mnras, 229, 423
\bibitem[Dressler \& Gunn(1983)]{dressler83}
Dressler, A. \& Gunn, J. E. 1983, \apj, 270, 7
\bibitem[Dressler et al.(2009)]{dressler09}
Dressler, A., Rigby, J., Oemler, A., Fritz, J., Poggianti, B. M., Rieke, G. \& Bai, L.
2009, \apj, 693, 140
\bibitem[Duc et al.(2002)]{duc02}
Duc, P. A. et al. 2002, \aap, 382, 60
\bibitem[Eisenstein et al.(2006)]{eisenstein06}
	Eisenstein, D. J., et al., 2006, \apjs, 167, 40
\bibitem[Ferrarese \& Merritt(2000)]{ferrarese00}
Ferrarese, L. \& Merritt, D. J. 2000, \apj, 539, L9
\bibitem[Georgakakis et al.(2008)]{georgakakis08}
Georgakakis, A. et al. 2008, \mnras, 406, 420
\bibitem[Goto et al.(2003)]{goto03} Goto, T. et al. 2003, \pasj, 55, 771
\bibitem[Goto(2004)]{goto04} Goto, T. 2004, \aap, 427, 125
\bibitem[Goto(2005)]{goto05} Goto, T. 2005, \mnras, 357, 937
\bibitem[Goto(2007)]{goto07}
Goto, T. 2007, \mnras, 381, 187
\bibitem[Goto et al.(2008)]{goto2008}
Goto, T. , Yagi, M. \& Yamauchi, C., 2008, \mnras, 391, 700
\bibitem[Haines et al.(2009)]{haines09}
Haines, C. P. et al. 2009, \apj, 704, 126
\bibitem[Hernquist(1989)]{hernquist92} Hernquist, L. 1989, \nat, 340, 687
\bibitem[Hopkins et al.(2008)]{hopkins08}Hopkins, P. F. et al. 2008, \apjs, 175, 390
\bibitem[Kennicutt(1998)]{kennicutt98} Kennicutt, R. C. 1998, \araa, 36, 189
\bibitem[Liu et al.(2007)]{liu07}
Liu, C. T., Hooper, E. J., O'Neil, K., Thompson, D., Wolf, M. \& Lisker, T. 2007, \apj, 658, 249
\bibitem[Melnick \& de Propris(2012)]{melnick12}Melnick \& de Propris, \mnras, {\sl in preparation} 
\bibitem[Miller \& Owen(2001)]{miller01}
Miller, N. A. \& Owen, F. N. 2001, \apj, 554, L25
\bibitem[Martini(2004)]{martini04}
Martini, P., 2004,  Coevolution of Black Holes and Galaxies, 169
\bibitem[Morrison et al.(2003)]{morrison03}
Morrison, G. E. et al. 2003, \apjs, 146, 267
\bibitem[Panuzzo et al.(2007)]{panuzzo07}
Panuzzo, P. et al. 2007, \apj, 656, 206
\bibitem[Parma et al.(2007)]{parma07}
Parma, P., Murgia, M., de Ruiter, H. R., Fanti, R., Mack, K.-H. \& Govoni, F.
2007, \aap, 470, 875
\bibitem[Poggianti et al.(1999)]{poggianti99}
Poggianti, B. M. et al. 1999, \apj, 518, 576
\bibitem[Poggianti \& Wu(2000)]{poggianti00}
Poggianti, B. M. \& Wu, H. 2000, \apj, 529, 157
\bibitem[Sadler et al.(2002)]{sadler02}
Sadler, E. M. et al. 2002, \mnras, 329, 227
\bibitem[Saintonge et al.(2008)]{saintonge08}
Saintonge, A., Tran, K.-V. H. \& Holden, B. P. 2008, \apj, 685, L113
\bibitem[Shin et al.(2011)]{shin11} Shin, M.-S., Strauss, M. A. \& Tojeiro, R. 2011, \mnras, 410, 1583
\bibitem[Smail et al.(1999)]{smail99} Smail, I. et al. 1999, \apj, 525, 609
\bibitem[Shioya et al.(2002)]{shioya02}
Shioya, Y., Bekki, K., Couch W. J., De Propris, R. 2002, \apj, 565, 223
\bibitem[Snyder et al.(2011)]{snyder11}Snyder et al. 2011, \apj, 741, 77
\bibitem[Vergani et al.(2010)]{vergani10}Vergani, D. et al. 2010, \aap, 509, 42
\bibitem[White et al.(2007)]{white07}
	White, R. L., Helfand D.J., Becker R.H., Glikman E. \& de Vries W., 2007, \apj, 654, 99
\bibitem[Yagi et al.(2006)]{yagi2006}
Yagi, M., Goto, T. \& Hattori., T., 2006 \apj, 624, 152
\bibitem[Yang et al.(2004)]{yang04}
Yang, Y., Zabludoff, A., Zaritsky, D., Lauer, T. R. \& Mihos, C. 2004, \apj, 607, 258
\bibitem[Yang et al.(2008)]{yang08}
Yang, Y., Zabludoff, A., Zaritsky, D. \& Mihos, C. 2008, \apj, 688, 965
\bibitem[York et al.(2000)]{york00} York, D. G. et al. 2000, \aj, 120, 1579
\bibitem[Yun et al.(2001)]{yun01}
Yun, M. S., Reddy, N. A. \& Condon, J. J. 2001, \apj, 554, 803
\bibitem[Zabludoff et al.(1996)]{zabludoff96}
Zabludoff, A. I. et al. 1996, \apj, 466, 104
\end{thebibliography}
\end{document}